\documentclass[twocolumn,aps,showpacs]{revtex4}
\usepackage{graphicx}
\def\PRL{Phys. Rev. Lett. }
\def\PRC{Phys. Rev. C}

\def\PLB{Phys. Lett. B}

\def\etal{\emph{et al.}}
\def\pT{$p_T$}

\def\Nbinary{\mbox{$\mathrm{N}_\mathrm{bin}$}}
\def\NbinaryMean{\mbox{$\langle\Nbinary\rangle$}}
\def\sigmaNNinel{\mbox{$\sigma^{NN}_{inel}$}}
\def\TAA{\mbox{$T_{AA}$}}

\begin{document}

\title{Can hadronic rescattering explain the ``jet quenching" at RHIC?}

\author{David Hardtke}
\affiliation{Department of Physics \\ University of California \\ Berkeley, CA  94720}
\author{Thomas J. Humanic}
\affiliation{Department of Physics \\ The Ohio State University \\ Columbus, OH 43210}

\begin{abstract}
Recent RHIC data have shown novel nuclear modifications of moderate to high $p_T$ 
particle production in central Au+Au collisions, including a suppression of
hadron production and a disappearance of back-to-back hadron pairs.  
In this paper, we investigate
whether final-state hadronic interactions of the jet fragments can reproduce
the RHIC data.  We find that hadronic rescattering can account for the
disappearance of back-to-back hadron pairs, but cannot reproduce other
features of the RHIC data.  
\end{abstract}
\maketitle

Recent data on the production of high $p_T$ hadrons in central Au+Au collisions
at RHIC indicate novel nuclear effects.  Above $p_T = 5-6$ GeV/c, 
the yield of hadrons is suppressed
by a factor of $\approx 5$ compared to what would be expected from an 
incoherent superposition of inelastic nucleon-nucleon collision 
\cite{PHENIX_highpt,STAR_highpt}. Data on
two-particle azimuthal correlations show similar near-angle jet-like
correlations
in central Au+Au and p+p collisions \cite{starv2,STARbtob}.  
The back-to-back dihadrons 
indicative of dijet production, however, are absent or greatly suppressed
in the most central Au+Au collisions. In addition, the production
of high $p_T$ hadrons shows a strong azimuthal 
correlation with respect the reaction 
plane (``elliptic flow") \cite{starv2}.   

Taken together, these experimental data are thought to result 
from a novel nuclear effect known as jet quenching.  Recent
measurements from d+Au collisions do not show the same behavior as central
Au+Au collisions \cite{stardAu}, so the modification of moderate to high $p_T$ hadron
production observed in central Au+Au collisions are  
thought to arise primarily from the interaction
of fast partons or their fragmentation products with the dense medium produced
in central collisions of heavy nuclei.  The goal of the current work
is to investigate whether the data can be explained entirely in terms of 
the hadronic interactions of jet fragmentation products with a dense
hadronic medium.  

Fast partons traversing a dense gluonic medium are expected to lose energy
and acquire transverse momentum relative to their original direction of 
propagation \cite{Gyulassy,Wang,Baier}.  
This energy loss is due to radiative induced gluon emission.  The rate of
energy loss is proportional to the gluon density of the medium traversed. 
The previously mentioned RHIC data have been described quite successfully
by convoluting expected parton production rates, parton energy loss in 
an expanding dense medium, and parton fragmentation.  

To quantify the nuclear matter effects on
high $p_T$ particle production the nuclear modification factor 
$R_{AA}$ is constructed,
\begin{equation}
\label{RAA}
R_{AA}=\frac{d^2N^{AA}/d{p_T}d\eta}{\TAA{d}^2\sigma^{NN}/d{p_T}d{\eta}}\ ,
\end{equation}
\noindent
where the nuclear overlap function \TAA=\NbinaryMean/\sigmaNNinel\ from a 
Glauber calculation accounts for 
the nuclear collision geometry. 
In the absence of nuclear matter effects, $R_{AA}$ should approach unity at 
moderate $p_T$ (2-3 GeV/c). In central Au+Au collisions at RHIC, $R_{AA}
\approx 1/5$.  This behavior was predicted by models that incorporate partonic
energy loss in a dense gluonic medium.
A recent paper, however, pointed out that this suppression of single inclusive
particle production at high $p_T$ can also be explained qualitatively by 
assuming
that partons fragment inside a dense hadronic medium \cite{Gallmeister}.  
The basic feature
of the single inclusive data ($R_{AA} \approx 1/5$) is reproduced by this
model, although it may fail to describe the exact $p_T$ dependence of the
suppression.  Adopting the notion of a colorless pre-hadron allows for a better
description of the $p_T$ dependence of $R_{AA}$\cite{Cassing}. With suitable 
modifications to the initial state nuclear effects
it is likely that this hadronic rescattering model could give a reasonably
qualitative description of the single inclusive data, including the
exact $p_T$ dependence of $R_{AA}$.  

The ability to describe the single inclusive data both in terms of parton
energy loss in a dense gluonic medium and hadronic rescattering and energy loss
in a dense hadronic medium is not surprising.  Any many-body calculation
based on the Boltzmann equation yields identical single particle
distributions under the substitution 
$\sigma \rightarrow A\sigma$ and $\rho \rightarrow \rho/A$ where $\sigma$ is 
the
two-body scattering cross-section, $\rho$ is the density of scattering 
centers, and $A$ is an arbitrary number.  Thus, a single particle observable
is unable to distinguish between a dense medium with small scattering 
cross-sections and a more dilute medium with larger scattering 
cross-sections\cite{humanic_subdivision}.
In general, QCD energy loss cannot be described using the Boltzmann equation
as QCD energy loss processes may be coherent.  
Nonetheless, the uncertainties in the
dynamical evolution of a heavy-ion collision coupled with the uncertainties
in nuclear effects in parton production from cold nuclei make it difficult
to distinguish between the partonic and hadronic energy loss scenarios using
single particle data alone.

Fluctuation observables are not invariant, however, under the substitution
$\sigma \rightarrow A\sigma$ and $\rho \rightarrow \rho/A$.  For $A>1$,
fluctuations will be increased.  Fluctuations can be measured via many 
observables, but are most easily quantified in terms of two particle 
correlations.  The production of jets in a heavy-ion collision can 
be thought of in terms of a local fluctuation in high $p_T$ particle production. 
In this paper, we investigate
what a hadronic rescattering interpretation of the RHIC single inclusive
particle production data would predict for two-particle azimuthal correlations.  We find that the hadronic rescattering
picture is unable to simultaneously describe the RHIC data on inclusive
particle suppression and
high-$p_T$ azimuthal correlations.  
The paper is organized as follows.  We first compare
RHIC directly to the PYTHIA event generator and find that this model reasonable
describes the features of moderate to high $p_T$ inclusive particle production
and two-particle azimuthal correlations.  We then introduce a hadronic
rescattering model that will be used to model a dense hadronic system into
which we will embed fragmentation products from PYTHIA events. We then merge
the PYTHIA jet events with the rescattering model, and study the propagation
of PYTHIA fragmentation products in our dense hadronic medium.  We find 
quantitative disagreement with the RHIC data, particularly due to the copious
resonance production from hadronic rescattering that should be 
manifest in the two-particle azimuthal correlations but is not observed
in the experimental data.  

\section{The PYTHIA event generator: Comparison to RHIC data}  

\begin{figure}[htb]
\begin{center}
\includegraphics[width=8cm]{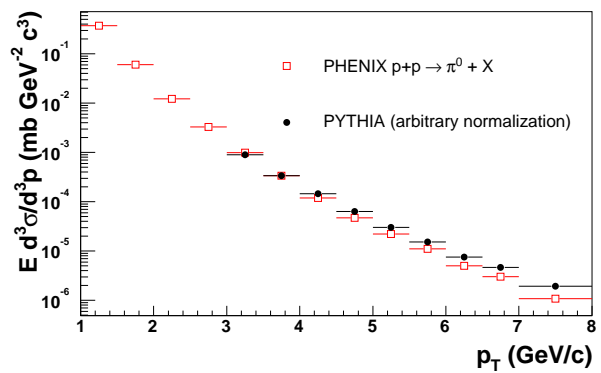}
\end{center}
\caption{Comparison of the invariant spectra from RHIC $p+p$ data \protect{\cite{PHENIXpp}} and PYTHIA.  The
PYTHIA spectrum is arbitrarily normalized.}
\label{ppspectra}
\end{figure}

In order to model hard scattering processes and fragmentation, we use the
PYTHIA event generator \cite{lund}.  This model convolutes measured parton
distribution functions, elementary parton-parton scattering cross-sections
and a phenomenological model of jet fragmentation.  In Figure \ref{ppspectra}
we compare the invariant $p_T$ spectra for pions produced running PYTHIA with its 
standard settings with the recently measured PHENIX $p+p$ data \cite{PHENIXpp}.  
For our studies
here we are only interested that the shape of the PYTHIA spectra is similar
to the data, so the PYTHIA spectrum is normalized to match the real data 
at $p_T = 4$ GeV/c.  As seen, the PYTHIA event generator with default settings 
does a reasonable
job in describing the measured RHIC data between 4 and 8 GeV/c.

\begin{figure}[htb]
\begin{center}
\includegraphics[width=8cm]{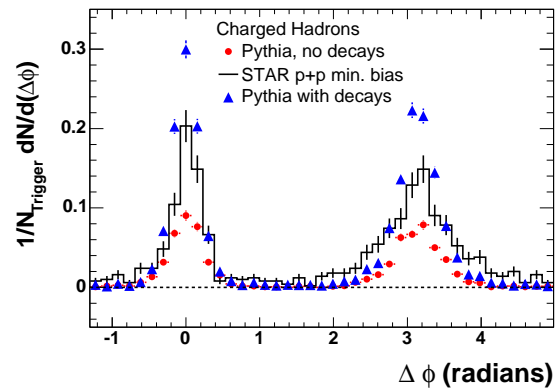}
\end{center}
\caption{Per trigger hadron relative azimuthal distributions for STAR data \protect{\cite{STARbtob}} compared to PYTHIA calculation with and without resonance decays.}
\label{C2decay}
\end{figure}

Our main goal in this paper is to investigate azimuthal correlations among
high $p_T$ hadrons. In Figure \ref{C2decay} we compare the azimuthal
correlations measured by the STAR collaboration \cite{STARbtob} to those produced by the PYTHIA
event generator.  The PYTHIA calculations use identical kinematic cuts as
the STAR collaboration (pseudorapidity $\eta<0.7$).  Events with a 
high $p_T$ trigger hadron with $4<p_T<6$ GeV/c are found.  We then 
calculate the azimuthal separation of other hadrons
in these events with $p_T>2$ GeV/c.  The resulting azimuthal distribution
is normalized to the number of trigger hadrons.  In both the STAR data and
the PYTHIA calculations there are strong azimuthal correlations near 
$\Delta\phi\approx0$ and $\Delta\phi\approx\pi$.  The correlated back-to-back
hadrons arise from the fragmentation of back-to-back dijets.   

The correlated small-angle pairs can be produced via two mechanisms.  During
jet fragmentation, the produced hadrons will be collimated thereby producing
azimuthally correlated particle pairs.  Such correlated pairs, however, can
also be the result of the decay of resonances.  
In Figure \ref{C2decay} we investigate the role of resonance decays
in the PYTHIA calculations.  Within this model, 
stable hadrons ($\pi,K,p$) are produced
directly during string fragmentation and also during the decay of 
unstable resonances produced during string fragmentation.  
In Figure \ref{C2decay}
we show two different PYTHIA calculations.  The standard PYTHIA setting includes
decay of unstable resonances and we see a strong near-angle and back-to-back
correlation peak.  We also ran PYTHIA in a mode where no resonances produced 
during fragmentation were allowed to decay.  In this mode near-angle and
back-to-back correlations are also observed.  These correlations, however,
are greatly suppressed.  It is important to note, however, that the suppression
factor is identical for the near-angle and back-to-back hadron pairs.  Since
the back-to-back hadron pairs cannot arise from low mass resonance 
decay (the invariant mass of the parent would need to be 
greater than 5 GeV) we conclude
that the near-angle correlations observed in PYTHIA arise primarily from
the jet fragmentation and are not due to resonance decay.  Turning
on the resonance decays leads to the observation of more charged hadrons per
event and hence increases the strength of the near-angle and back-to-back
azimuthal correlations.  Analogously, if we were to include $\pi^0$ in the
construction of the azimuthal distribution, we would see a \emph{trivial} 50\%
increase in the near-angle and back-to-back correlation strength.  The 
increase of the azimuthal distributions is due primarily to the observation of the extra
hadrons from resonance decay and not from the correlations induced by these
decays. 

It should be noted that the absolute per-trigger yield of associated particles
differs between PYTHIA and the STAR data.  In the case of PYTHIA with decays,
the per-trigger yield of associated particles is larger for PYTHIA compared 
to the STAR data. In the case of PYTHIA without decays, the per-trigger yield
of associated particles is smaller than the STAR data.  No attempt was made
to adjust the PYTHIA fragmentation settings to rectify this difference.  
These azimuthal distributions are normalized to the number of 
``trigger hadrons", and these mostly arise from the case where the leading
hadron carries a large fraction of the parton momentum.  This is the region
where the experimental data on parton fragmentation functions is most 
uncertain, and PYTHIA fragmentation functions are tuned to experimental data.  
Thus any discrepancy between measured dihadron correlation data 
and the PYTHIA calculations is not of fundamental concern.     

\section{A Simple Hadronic Rescattering Model}

The hadronic rescattering model \cite{humanic} that we will use to understand
high $p_T$ hadron propagation in a dense hadronic medium 
has been described elsewhere
and was shown to reproduce many of the low $p_T$ experimental observables at 
RHIC
including transverse momentum distributions for $\pi, K, p$, $v_2(p_T)$, and
the Hanbury-Brown Twiss radii\cite{humanic_preprint}. 
We briefly review some of the features of
the model below.

Rescattering is simulated with a semi-classical 
Monte Carlo calculation which assumes strong binary collisions 
between isospin-averaged hadrons. Relativistic kinematics is used 
throughout.  All calculations are made to simulate RHIC-energy
Au+Au collisions in order to compare with the results of
RHIC data.

The initial stage of the rescattering calculation 
employs simple parameterizations 
to describe the 
initial momenta and space-time of the hadrons. The initial 
momenta are assumed to follow a thermal-like transverse
momentum distribution for all particles,
\begin{equation}
(1/{m_T})dN/d{m_T}=C{m_T}/[\exp{({m_T}/T)} \pm 1]
\end{equation}
where $T$ is a ``temperature parameter'', 
and a Gaussian rapidity distribution for mesons,
\begin{equation}
dN/dy=D \exp{[-{(y-y_0)}^2/(2{\sigma_y}^2)]}
\end{equation}
where $\sigma_y$ is the rapidity width.
Two rapidity distributions for baryons have been tried: 1) flat
and then falling off near beam rapidity and 2) peaked at central
rapidity and falling off until beam rapidity. Both baryon
distributions give about the same results. 
The initial longitudinal
particle hadronization position ($z_{had}$) and time ($t_{had}$) 
are determined by the relativistic equations,
\begin{equation}
z_{had}=\tau_{had}\sinh{y};
\qquad  t_{had}=\tau_{had}\cosh{y}
\end{equation}
where $\tau_{had}$ is the 
hadronization proper time. From Equations 2 and 3, it is seen
that longitudinal invariance is \emph{not} assumed in the initial conditions 
for the present calculations. Calculations were carried
out using isospin-summed events containing at
freeze-out about 5000 pions, 500 kaons, and 650 nucleons
($\Lambda$'s were decayed).
The hadronization model parameters which were found to reproduce
the RHIC data were $T=300$ MeV,
$\sigma_y$=2.4, and $\tau_{had}$=1 fm/c.

\begin{figure}[htb]
\begin{center}
\includegraphics[width=8cm]{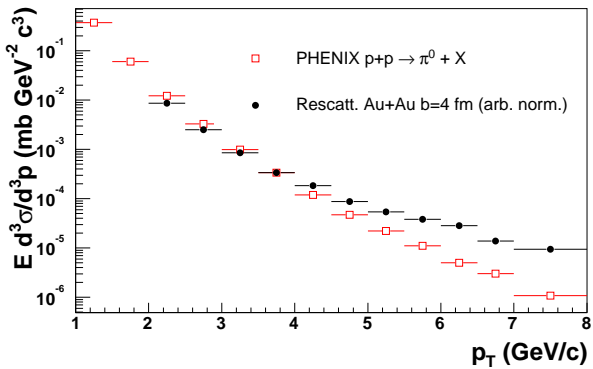}
\end{center}
\caption{Comparison of the invariant spectra from RHIC $p+p$ data and a 
pure rescattering model calculation for Au+Au collisions with b=4 fm.  The
rescattering model spectra is arbitrarily normalized.}
\label{rescattspectra}
\end{figure}

We now compare RHIC high $p_T$ data to calculations from the pure rescattering
model.  In other words, for these calculations 
hard scattering and fragmentation
is not implemented.  In Figure \ref{rescattspectra} we compare the 
PHENIX $p+p\rightarrow \pi^0 + X$ data to the pure rescattering model calculations for
Au+Au collisions with 
impact parameter $b=4$ fm.  For high $p_T$ $\pi^0$ production at RHIC, 
the spectral shape is similar in central Au+Au and p+p collisions. 
We are only interested in spectral shapes, so
the rescattering model calculations are arbitrarily normalized to the PHENIX
data at $p_T = 4$ GeV/c (The integral of the $p_T$ spectra and hence the total
yield is used to adjust the initial conditions of the calculation).  We see that the rescattering model produces copious high $p_T$ hadrons and exhibits
a power law behavior.  The rescattering model over-predicts the yield
of high $p_T$ hadrons.  This is probably due to the treatment of the densest
stage of the collision in terms of $2\rightarrow2$ binary hadronic collisions 
whereas a more correct treatment would model this stage using many-body
collision dynamics. It has been verified numerically that this approximation
does not affect the bulk dynamics of the collision 
\cite{humanic_subdivision} as measured by low $p_T$ observables. 

\begin{figure}[htb]
\begin{center}
\includegraphics[width=8cm]{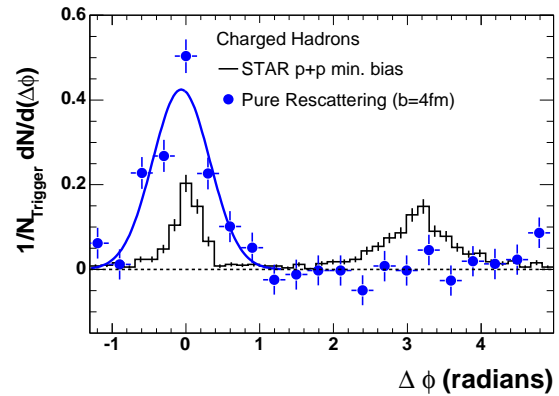}
\end{center}
\caption{Per trigger hadron relative azimuthal distributions for STAR $p+p$ 
data compared to a pure rescattering model calculation for $Au+Au$ at $b=4$ fm. 
The blue line shows a Gaussian
fit to the azimuthal distribution from the rescattering model.}
\label{C2rescatt}
\end{figure}

The pure rescattering model calculations also show near-angle 
(but not back-to-back) azimuthal correlations.  
Figure \ref{C2rescatt} compares the measured
azimuthal distributions for p+p collisions \cite{STARbtob} 
and the azimuthal correlations
produced in the rescattering model calculation for $Au+Au$ at $b=4$ fm.  
The near-angle correlations
in the rescattering model come from the decay of resonances.  We have confirmed
this by considering rescattering particles that do not come directly from a 
resonance decay, in which case no azimuthal correlations are observed.  
These correlations are both stronger and broader than those measured in 
the p+p data. We have fit the rescattering model near-angle azimuthal 
distribution with a Gaussian and get a width of $\sigma = 0.39 \pm 0.04$.  In contrast
STAR has measured the widths of the near-angle azimuthal correlation in p+p, 
d+Au, and central Au+Au collisions and gets a width in all cases of 
$\sigma\approx 0.2$ for these kinematic cuts. 
The large discrepancy in the widths of these correlations
indicates a fundamentally different origin.  In the rescattering model, the
correlations come from resonance decay, whereas for the RHIC data the 
correlations seem to arise primarily from jet fragmentation.    

The near-angle azimuthal correlation strength is related
to the yield of resonances at $p_T \approx p_T (trigger) + p_T (associated)$.  The
fact that the pure rescattering model produces an invariant $p_T$ spectrum
of hadrons and resonances that is harder than that measured in the data leads
to a somewhat artificial increase in the near-angle correlation strength.
This model is certainly incorrect for calculating moderate
to high $p_T$ particle production in Au+Au collision at RHIC, so we chose 
to make no attempt to 
correct the model
for its possible shortcomings.  The rescattering model will only be 
used as a rough
estimate of the environment that a jet fragmentation product would
see if it was able to fragment inside such a dense hadronic medium.
Nonetheless, we note that this pure rescattering model does
produce azimuthal correlated particle pairs due to resonance decay, and that
these correlations are broader and stronger than those seen in the real data.

\section{The Propagation of PYTHIA Test Particles in the Rescattering Model} 

PYTHIA events are embedded into these rescattering events, with the production
vertices chosen to account for the initial nuclear overlap geometry.  
The embedded PYTHIA events are required to have a high $p_T$ hadron with
\pT $>$ 3 GeV/c.  The particles coming from these PYTHIA events are tagged
as such, and any resonance produced via the interaction of a PYTHIA particle
is tagged as a ``PYTHIA resonance".  The decay products of these PYTHIA 
resonances are tagged as PYTHIA particles, and in the plots that follow we
only look at the final-state PYTHIA particles.  Thus, the artificially large
high $p_T$ particle production seen in the pure rescattering model does not
invalidate our subsequent calculations.  

The main theoretical uncertainty in these studies is the space-time
development of hadrons in parton fragmentation.  In the standard PYTHIA
fragmentation scheme, no attempt is made to describe the space time aspects
of jet fragmentation.  There are many theoretical models of how jet
fragmentation develops,
but little experimental support for any single idea.  In the conventional
QCD model, hadrons from parton fragmentation are formed at time $t=ER^2$\cite{Dokshitzer}, where $E$ is the parton energy and $R$ is hadronic size.  Thus,
very high $p_T$ hadrons would necessarily form well outside the medium produced
in a Au+Au collision at RHIC.  In 
another picture colorless pre-hadrons are formed very early\cite{Kapeliovich} 
in which case hadrons would
form deep inside the medium produced at RHIC.  The pre-hadrons form instantly
in the limit where $z_h\rightarrow 1$, where $z_h$ is the fraction of the
parton momentum carried by the leading hadron.    
One argument against such a 
hadron formation picture at RHIC is that hadrons should not exist in a
quark-gluon plasma. The instantaneous formation of a colorless pre-hadron is
based on the notion of vacuum energy loss and motivated by the string
picture.  In a quark-gluon plasma, Debye screening leads to the modification
of the string tension and would reduce the rate of vacuum energy loss.

Here we adopt
a naive picture of hadron formation, investigate its consequences, and rule it
out along with any similar fragmentation picture.
In these calculations, the hadrons from the PYTHIA events initially form at proper
time, $\tau$, and at (z,t) as given by Equation 3 with $\tau_{had}=\tau=1$ fm/c,
to be consistent with the rest of the calculation. In the transverse direction, the
(x,y) position of the embedded PYTHIA events are 
randomly selected using the nuclear
overlap model.  The individual hadrons in each PYTHIA event are further  
smeared in a circle of radius
1.5 fm. This radius is determined by the condition that it is large enough 
such that hadronic rescattering within a PYTHIA event is small.

\begin{figure}[htb]
\begin{center}
\includegraphics[width=8cm]{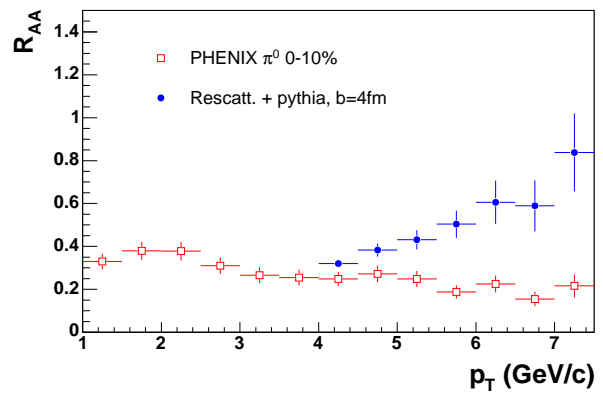}
\end{center}
\caption{Nuclear Modification Factor $R_{AA}$ for pions.  The open squares show
PHENIX $Au+Au$ data for the 10\% most central collisions, and the solid circles show
the results of the rescattering model with embedded PYTHIA events.  Only particles 
tagged as originating from PYTHIA or a PYTHIA-particle induced resonance 
are included in the rescattering+PYTHIA calculation.}
\label{RAAfig}
\end{figure}

Figure \ref{RAAfig} shows the nuclear modification factor $R_{AA}$ for pions
from these calculations as well as PHENIX $Au+Au$ data \cite{PHENIX_highpt} at a 
similar impact parameter.
For the model calculations, $R_{AA}$ is defined as:
\begin{equation}
R_{AA}(p_T) = \frac{N(\mathrm{PYTHIA\;tagged\; from\; 
rescattering\; events})}{N(\mathrm{\;PYTHIA\;p+p})}.
\end{equation}   
With this definition, we consider only the modification of the spectra of
the embedded PYTHIA events, and do not consider the contribution to particle
production coming from the rescattering model.  Because we inject triggered
PYTHIA events into the rescattering model, 
we only consider $R_{AA}$ for $p_T>4$ GeV/c.

Figure \ref{RAAfig} shows similar features to the calculations of 
Gallmeister et al.\cite{Gallmeister}.  The absolute value of $R_{AA}$ is 
well reproduced
in the moderate $p_T$ region ($p_T \approx 4$ GeV/c), but $R_{AA}$ trends 
upwards with increasing $p_T$, a trend not seen in the experimental data.
The $p_T$ dependence arises from our treatment of the space-time development
of hadrons in jet fragmentation.  Higher $p_T$ hadrons are not able to interact
until proper time $\tau$ and they thus travel a distance of $\gamma \tau$ 
in the laboratory frame (here $\gamma$ is the Lorentz factor).

\begin{figure}[htb]
\begin{center}
\includegraphics[width=8cm]{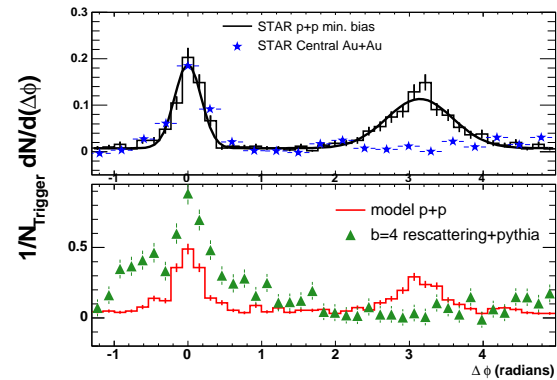}
\end{center}
\caption{Per trigger hadron relative azimuthal distributions for 
STAR data compared to PYTHIA+rescattering model calculations.  
Only particles tagged as originating from PYTHIA or a PYTHIA-particle 
induced resonance are include in the rescattering+PYTHIA calculation.}
\label{C2rescattWithPythia}
\end{figure}

Figure \ref{C2rescattWithPythia} shows the two-particle 
azimuthal distributions from the STAR
collaboration \cite{STARbtob} and from the rescattering events with embedded
PYTHIA jets.    For the model calculations, we only look at hadrons
produced directly from the PYTHIA event or produced via the decay of
a PYTHIA particle induced resonances.  In the
model calculations all hadrons are considered (charged and neutral)
whereas for the STAR measurements only charged hadrons are used.  This
should lead to $\approx 50\%$ stronger correlations in the model compared to
the real data.  

As pointed out by STAR, the correlations near $\Delta\phi\approx0$ are 
indicative of jet fragmentation, and are very similar in magnitude and width
for proton-proton and central Au+Au collisions.  The azimuthally back-to-back
dihadron pairs ($\Delta\phi\approx\pi$), 
indicative of dijet production, are present in the 
proton-proton collisions but absent in the most central Au+Au collisions.  
This feature is well reproduced in the rescattering model + PYTHIA jet 
calculations.  

Focusing, however, on the region $\Delta\phi\approx0$ reveals a striking 
difference between the rescattering+PYTHIA calculations and the STAR data.
While the STAR data shows little or no modification of the correlation 
structure in central Au+Au collisions compared to the proton-proton reference,
the rescattering model leads to a broadening and enhancement of
the near-angle correlations.  
This is exactly the structure observed in the pure rescattering model 
calculations (where no PYTHIA jets were embedded into the events).  
These correlations are due primarily to the resonances produced
during the rescattering stages of the collisions, and do not come from the 
initial jet fragmentation.  We thus conclude that the near-angle correlation
structure observed by STAR in central Au+Au collision is due to jets that
fragment outside the medium.  If the jets were to fragment inside a dense
hadronic medium they would produce resonances that lead to broad azimuthal
correlations.  The broad resonance induced correlations seen in the 
PYTHIA+rescattering calculations are not seen in the central STAR Au+Au data.

\section{Conclusions}

The failure of this model to describe in detail all aspects of high $p_T$ 
particle production at RHIC is not surprising.  It is unlikely that a 
$2\rightarrow 2$
hadronic scattering description is valid in the early stages of a RHIC 
collision.  For low $p_T$ observables such as elliptic flow and HBT this model
does seem to work.  At high $p_T$, however, we show that this model does not
work and we feel that no purely hadronic model can describe the 
``jet quenching" observables seen in central RHIC collisions.  
The nuclear suppression factor
will have an unwanted $p_T$ dependence.  In addition, the unavoidable copious 
resonance production will lead substantial modification in the two-particle 
high $p_T$ azimuthal correlations. In particular, we find that resonance decays
lead to broader high $p_T$ azimuthal correlations, and there is no evidence
in the RHIC data for large broadening of the azimuthal correlations.  In 
general, we find that only correlation and fluctuation measurements, of 
which high $p_T$ azimuthal correlations are one example, are able
to distinguish between a purely hadronic description of RHIC data and an
interpretation in terms of novel forms of nuclear matter such as the 
quark-gluon plasma.  

\section{Acknowledgments}
We would like to thank Michael Lisa for help with the manuscript and useful
discussions.  We thank the staff of the Ohio Supercomputing Center, where
these calculations were made.  This work was supported by the National Science
Foundation under grants PHY-0099476 and PHY-0203111.

\bibliographystyle{unsrt}

\end{document}